\documentclass[pdflatex,sn-nature]{sn-jnl}

\usepackage{graphicx}%
\usepackage{siunitx}
\usepackage{multirow}%
\usepackage{amsmath,amssymb,amsfonts}%
\usepackage{amsthm}%
\usepackage{mathrsfs}%
\usepackage[title]{appendix}%
\usepackage{xcolor}%
\usepackage{textcomp}%
\usepackage{manyfoot}%
\usepackage{booktabs}%
\usepackage{algorithm}%
\usepackage{algorithmicx}%
\usepackage{algpseudocode}%
\usepackage{listings}%
\usepackage{pdfpages}%

\theoremstyle{thmstyleone}%
\theoremstyle{thmstyletwo}%

\theoremstyle{thmstylethree}%

\begin{document}

\title{A Collaborative Framework Integrating Large Language Model and Chemical Fragment Space: Mutual Inspiration for Lead Design}


\author[1,2]{\fnm{Hao} \sur{Tuo}}

\author[1,2]{\fnm{Yan} \sur{Li}}

\author[1,2]{\fnm{Xuanning} \sur{Hu}}

\author*[3]{\fnm{Haishi} \sur{Zhao}} \email{zhaohs@jlu.edu.cn}
\author*[1,2]{\fnm{Xueyan} \sur{Liu}} \email{lxy@jlu.edu.cn}
\author*[1,2]{\fnm{Bo} \sur{Yang}} \email{ybo@jlu.edu.cn}

\affil[1]{\orgdiv{College of Computer Science and Technology}, \orgname{Jilin University}, \city{Changchun}, \country{China}}

\affil[2]{\orgdiv{Key Laboratory of Symbolic Computation and Knowledge Engineering of Ministry of Education}, \orgname{Jilin University}, \city{Changchun}, \country{China}}

\affil[3]{\orgdiv{College of Earth Sciences}, \orgname{Jilin University}, \city{Changchun}, \country{China}}

\maketitle
\section*{Abstract}
Drug design, particularly in the discovery of lead compounds, is of core strategic importance to combating disease and enhancing human well-being. Prevailing computational methods, however, struggle to effectively integrate domain-specific knowledge, severely limiting their capacity to identify novel lead compounds with validated binding modes and new scaffolds.
Here, we propose AutoLeadDesign, a lead compounds design framework that inspires extensive domain knowledge encoded in large language models with chemical fragments to progressively implement efficient exploration of vast chemical space. The comprehensive experiments indicate that AutoLeadDesign outperforms baseline methods. Significantly, empirical lead design campaigns targeting two clinically relevant targets (PRMT5 and SARS-CoV-2 PLpro) demonstrate AutoLeadDesign's competence in de novo generation of lead compounds, achieving expert-competitive design efficacy. Structural analysis further confirms their mechanism-validated inhibitory patterns. By tracing the process of design, we find that AutoLeadDesign shares analogous mechanisms with fragment-based drug design, which traditionally rely on expert decision-making, further revealing why it works. Overall, AutoLeadDesign offers an efficient approach for lead compound design, suggesting its potential utility in drug design.
\newpage
\section{Introduction}

Drug design represents a cornerstone in confronting global health challenges and advancing sustainable societal progress. The core focus of this field lies in systematically exploring the chemical space to identify bioactive compounds with high affinity to target proteins, alongside novel interaction modalities capable of achieving breakthrough therapeutic effects\cite{1,2,17,18,19,20,21,22,3,4,5,6,7,8}.
Traditional approaches\cite{trad} rely heavily on expert decision-making, and this dependence on potentially biased intuition often results in limited exploration of chemical space, ultimately constraining the efficiency of drug design. Meanwhile, generative models\cite{tamgen,diff2} are trained to learn the distribution of known active compounds, which discourages finding novel molecules and leads to unsatisfactory optimization performance.

A more straightforward solution is combinatorial optimization algorithm that searches the discrete chemical space oriented by affinity\cite{survy1,JTVAER,autogrow4,reinforega,reinvent,smilesga}. Starting from random compound sets, these methods are particularly valuable for designing drugs targeting novel proteins lacking known bioactive inhibitors. However, existing methods face a critical challenge in fully modeling domain-specific knowledge to guide the process of design, struggling to identify bioactive compounds.

The large language models (LLMs) advancing in comprehensively comprehending, summarizing and utilizing knowledge, have demonstrated remarkable capabilities in handling domain-specific tasks\cite{37,38,39}. Recent studies\cite{LLM1,LLM2,LLM3,LLM4,LLM5,LLM6,LLM7,LLM8,LLM9,LLM10} reveal that LLMs now exhibit chemical reasoning capabilities comparable to human experts. Notably, emerging evidence\cite{LLM11} suggests LLMs may soon attain Turing test-level performance. All these findings indicate that LLMs have implicitly internalized substantial domain knowledge. Hence, a critical research frontier lies in developing methodologies to effectively harness latent knowledge of LLMs for solving real-world problems.
The most recent property-oriented molecular design model incorporating LLM is LMLF\cite{46}, which utilizes LLMs to iteratively optimize high-potential molecules for exploring uncharted chemical space.
However, experimental results show that compounds designed by LMLF exhibit pronounced tendency toward local optimization, demonstrating limited capacity to explore uncharted chemical space (Fig. 2a). This presents substantial challenges to de novo drug design against targets lacking known bioactive compounds, particularly for rare diseases or newly identified pathogenic proteins. This limitation results from the vastness of chemical space and the strict constraints for LLM.
Fragment-based design which utilizes molecular fragments, rather than individual atoms, as fundamental building blocks for constructing novel drug, owing to its significant reduction of search space and enhanced structural flexibility in molecular architecture, presents a promising alternative for improving the performance of LLMs in lead compound design.

In this study, we propose AutoLeadDesign, a framework that integrates LLM and chemical fragments to efficiently explore the chemical space for lead compounds design. The closed-loop workflow is illustrated in Fig. 1, containing two key mapping. The mapping from molecule space to fragment space (Fig. 1b) decomposes explored molecules into fragments to maximize the freedom of design and the quantitative analysis of occurrence frequency and contribution to binding energy updates the understanding of target-relevant chemical space. The mapping from fragment space to molecule space leverages rich domain knowledge embedded in LLM to further explore chemical space on the basis of current cognition. The process executes iteratively facilitating the sufficient inspiration to LLM and the comprehensive exploration of the chemical space, which is essential to discover novel and high-affinity lead compounds to target protein.


Notably, AutoLeadDesign demonstrated superior performance compared to other baseline models in de novo drug design. Through two case studies targeting PRMT5 and PLpro, we demonstrate the capability of AutoLeadDesign to de novo generate lead compounds exhibiting high binding affinity and unprecedented binding mechanisms, independently on the guide of known bioactive ligands. The designed compounds demonstrating superior docking score and the binding mode can be confirmed valid by related research, indicating highly bioactive. Furthermore, multidimensional analysis of the design process reveals the underlying mechanisms of fragment growing, merging, and linking, which align with conventional fragment-based drug design (FBDD) strategies, thereby elucidating the rationale for its effective functionality. These findings highlight the broad applicability and considerable potential of AutoLeadDesign in drug design.

\section{Results}


\subsection{AutoLeadDesign outperforms baseline methods in de novo molecular design}


To benchmark the overall performance of AutoLeadDesign, we compare our method against three combinatorial optimization methods: REINVENT\cite{reinvent}, ChemGE\cite{smilesga}, RGA\cite{reinforega} and an LLM-based method, LMLF\cite{46}. Commencing with randomly selected compound libraries, these methodologies progressively optimize molecular designs through iterative evaluation and expansion of uncharted chemical space. We randomly select 10 protein binding pockets in CrossDocked2020 dataset\cite{crossdocked} as test set to evaluate these methods and AutoLeadGen. To ensure the comprehensiveness of validation, these proteins correspond to distinct classes: hydrolase, transferase, oxidoreductase, member protein and immune system protein. In fact, the initial compound set may significantly impact the direction of exploration and the final result. To enable comprehensive evaluation, we have established two distinct initial compound sets under different experimental scenarios. One is random initialization: we randomly select 100 molecules from ZINC as 0-th generation. Another is prior initialization: on the basis of random set, we add the co-crystal ligand of target proteins in CrossDocked2020 dataset. We conducted 20 generations of optimization cycles, with each generation involving the design of 100 compounds per target protein in the test set. In practice, pharmacologists concern about the high-affinity compounds rather than the overall distribution of designed compounds. Hence, we select top-100 compounds with the best docking scores for evaluation and consider the TOP-1/10/100 metrics (average docking score of top-1/10/100).
\begin{table}[htbp]
	\centering
	\caption{\centering{Result of the de no design task in the scene of random and prior initialization}}
	\begin{tabular}{*{7}{c}}
		\toprule
		\multirow{2}*{Methods} & \multicolumn{3}{c}{Random} & \multicolumn{3}{c}{Prior} \\
		\cmidrule(lr){2-4}\cmidrule(lr){5-7}
		& Top-1 & Top-10 & Top-100 & Top-1 & Top-10 &Top-100 \\
		\midrule
		ChemGE & -10.68 & \underline{-10.35} & \underline{-9.82} & -10.71 & -10.36 & \underline{-9.85} \\
		RGA&-7.57&-7.03&-6.15&-9.12&-8.73&-7.02\\
		Reinvent & \underline{-10.73} & -10.2 & -9.49 & -10.58 & -10.11 &-9.48 \\
		LMLF & -9.20 &-8.70 & -7.59 & \underline{-10.96} & \underline{-10.41} & -8.89 \\
		AutoLeadDesign & \textbf{-11.51} & \textbf{-10.69} & \textbf{-9.85} & \textbf{-11.73} & \textbf{-11.15} & \textbf{-10.33} \\
		\bottomrule
	\end{tabular}
\end{table}

As shown in Table 1, AutoLeadDesign demonstrates statistically superior performance in both random and prior initialization scenarios, designing lead compounds with highest binding affinity compared to baseline methods. In prior initialization scenarios, AutoLeadDesign and another LLM-based method, LMLF, achieved optimal and sub-optimal performance, respectively, illustrating the effectiveness of LLM in drug design.
In the random initialization scenario, AutoLeadDesign remains the SOTA performance, demonstrating the capability in designing high-affinity compounds without reliance on bioactive compounds which are difficult to acquire. In contrast, the binding free energy of top compounds designed by LMLF exhibited a 16\% reduction (from -10.96 kcal/mol to -9.2 kcal/mol), indicating its sensitivity to initialization. This limitation may originate from the inherent characteristic of local optimization. In-depth analyses will be conduct in the following case study of PRMT5.

\subsection{Case study of PRMT5}

Protein Arginine Methyltransferase 5 (PRMT5) catalyzes the transfer of methyl groups from S-adenosylmethionine (SAM) to arginine residues forming symmetrical $\omega$-NG-dimethyl arginine on protein substrates, which plays a key role in crucial cellular pathways including cell growth and proliferation. The overexpression of PRMT5 involved in important cancer functions, indicating PRMT5 as a potential target for anticancer drug development\cite{PRMT51,PRMT52,PRMT53,PRMT54}. 
LLY is a high-affinity inhibitor targeting PRMT5, forming hydrogen bonds to ASP419 and MET420, which play crucial role in competing the binding pocket with SAM\cite{LLY1}.
In our lead design study, we initialized the process with LLY and 100 randomly selected ZINC compounds.

As shown in Fig. 2a, compounds designed by LMLF tend to cluster around the reference structure, demonstrating local optimization behavior. In contrast, AutoLeadDesign not only utilizes the known chemical space around LLY but also explores uncharted regions. In addition, Fig. 2b illustrates the binding patterns of compounds PRL001 and ADD001 designed by LMLF and AutoLeadDesign, respectively, with target protein. It can be seen that PRL001 exhibits a similar backbone and binding pattern to LLY. In contract, ADD001 has a novel structure. In the precept of binding interactions, ADD001 not only maintains interactions with key residues ASP419 and MET420 (similar to LLY), but also establishes an additional hydrogen bond with the key residue GLU444 in the substrate protein binding site\cite{PRMT53,PRMT54,JUN}. This dual binding mode represents a novel inhibition mechanism --- simultaneously competing with both SAM cofactor and protein substrate binding. The results illustrate the ability of AutoLeadDesign in designing high-affinity compounds with novel interaction mechanisms, distinct from merely optimizing known binding motifs.

To further evaluate the robustness of our approach, we construct the scenario of random initialization where we only randomly selected 100 compounds as start points. The visualization of results (Fig. 3a) shows that compounds designed by LMLF still cluster around initial structures, indicating similar low bioactivity. This characteristic of local optimization results in its sensitivity to initialization (Fig. 3b), thereby limiting the capability of LLMs to design lead compounds with novel binding modes to target proteins. By contrast, AutoLeadDesign exhibits exceptional proficiency in navigating uncharted chemical space, consistently yielding reliable results while demonstrating a key strength of low sensitivity to initial conditions. The analysis of binding interactions on top-ranked compounds --- PRR001 (Fig. 3c) and ADR001 (Fig. 3d) designed by LMLF and AutoLeadDesign respectively --- demonstrates that ADR001 forms a hydrogen bond with key residues GLU444 indicating high bioactivity while ADR001 only interacts with pharmacologically insignificant residues. These results underscore the importance of comprehensive chemical space exploration in lead compound design. The superior performance of AutoLeadDesign, which does not depend on difficult-to-obtain bioactive ligands, represents a significant advancement over LMLF, offering greater potential for novel drug design.

\subsection{Case study of PLpro}

The SARS-CoV-2 papain-like protease (PLpro), which mediates polyprotein processing, is indispensable for producing functional viral replicase complexes. Compared to existing antiviral therapeutics, PLpro-targeting agents may operate through divergent mechanisms of action, potentially providing improved therapeutic efficacy and resilience against emerging viral resistance\cite{PLpro1,PLpro2,PLpro3,PLpro4,PLpro5,PLpro6,PLpro7,PLpro8,PLpro9,PLpro10}. In this section, we exemplify the application of AutoLeadDesign in designing Plpro inhibitors in the setting of random initialization.

To validate the robustness of our methodology against variations in the initial compound library, we randomly constructed five molecular libraries as initial library.  The distribution of binding energy (Fig. 4a) demonstrates that AutoLeadDesign can robustly identify lead compounds (with a standard deviation of 0.16) with significantly higher binding affinity than known bioactive ligand ($\Delta G_{Jun1268}=$ $-9.6$ $kcal/mol$). As demonstrated in Figs. 4b-f, the generated molecules also exhibit favorable physicochemical properties. Notably, the majority of compounds display QED scores above 0.5, indicating excellent drug-likeness. The synthetic accessibility assessment reveals that most molecules possess SAScores below 5 (with a subset $\le$ 3), suggesting high synthetic feasibility. Furthermore, the calculated logP values fall within the pharmaceutically acceptable range, and molecular weights are predominantly distributed between 300-500 Da, consistent with typical small-molecule drugs. Importantly, over 78\% of the generated compounds comply with all the Lipinski five rules, underscoring their potential as drug candidates.

To further validate the rationality of results in binding interactions, we compare compounds designed by AutoLeadDesign (PLP001 and PLP002) with those designed by expert manually (GRL0617\cite{GRL1} and Jun12682\cite{Jun126821}). Both GRL0617 and Jun12682 form the $\pi-\pi$ stacking with TYR268 and hydrogen bonds with ASP164, and relative researches have validated the two interactions are essential for inhibitory activity\cite{GRL1,GRL2,GRL3,GRL4,Jun126821,Jun126822,Jun126823}. PLP001 (Fig. 5c) share similar binding mode with GRL0617 (Fig. 5a), both forming hydrogen bond and $\pi-\pi$ stacking with key residues ASP164 and TYR268 respectively, and the docking result shows that PLP001 have lower binding energy than GRL0617, indicating that PLP001 have stronger bioactivity than GRL0617. PLP002 also forming hydrogen bonds with key residues ASP164, meanwhile, it forms some bonds with neighbor residues, including $\pi-\pi$ stacking with TYR264, hydrogen bond with TYR273 and $\pi-cation$ with ARG166, which indicate a tight binding with pocket.
Overall, these novel results designed by AutoLeadDesign --- achieving expert-competitive performance --- demonstrate its exceptional capability in lead compound design.

\subsection{The chemical fragment space inspires LLM to make reasonable decision}

The superior performance of AutoLeadDesign primarily stems from its fragment-inspired and LLM-guided design paradigm, where chemical fragments inspire extensive domain knowledge embedded in LLM to guide the drug design process. Here, we conduct an in-depth analysis of the LLM decision-making process from the perspective of fragment-based drug design (FBDD) strategies (particularly fragment linking and merging strategies), demonstrating the scientific rationality of LLM-driven decisions within this workflow.

Fig. 6 illustrates the key decision-making process targeting PLpro. Fragment 1 and Fragment 2, derived from reference molecules Mol 1 and Mol 2 respectively (both exhibiting high binding potential), were selected as seed fragments and subsequently fed into LLM to generate Mol 3. Structural analysis revealed that the LLM designed a classic drug-like amide linker to bridge the two fragments. Detailed binding mode characterization demonstrated that both fragments occupy adjacent regions of the binding pocket and form $\pi-\pi$ stacking interactions with TYR-268 (Fig. 6b), exhibiting spatially proximal binding configurations. Notably, the amide linker not only preserves the native binding modes of the original fragments (Fig. 6d), but also forms an additional hydrogen bond with ASP-164 (Fig. 6c), significantly enhancing binding affinity (${\Delta\mathit{G}}_{Mol 1 \rightarrow 3}=$ $-1.8$ $kcal/mol$ and ${\Delta\mathit{G}}_{Mol 2 \rightarrow 3}=$ $-2.6$ $kcal/mol$). Moreover, this amide-containing fragment will enrich the fragment library to guide subsequent molecular design iterations. This workflow rigorously adheres to fundamental fragment linking principles, providing experimental validation for the scientific rationality of LLM-driven decision-making in fragment-based drug design (FBDD).

Fig. 7 demonstrates another pivotal decision-making process targeting PRMT5. Fragment 1 and Fragment 2 occupy opposite termini of the narrow binding pocket, interacting with LYS-393 and PHE-327 respectively. The overlap between the fragments precluded conventional linker design, prompting strategic replacement of the overlapping region with an amino group to achieve fragment merging. The resulting Mol 3 not only preserves the spatial orientations and binding modes of original fragments but also fully complies with the fundamental principles of fragment merging strategies.

In summary, detailed tracing of the decision-making process indicates that under the inspiration of sampled fragments, LLM intrinsically operationalizes the core tenets of fragment-based drug design (FBDD) with expert-like proficiency—mastering both fragment merging and linking strategies as canonical approaches.

\subsection{The exploration of fragment space}
AutoLeadDesign demonstrates comprehensive capabilities in molecular design by not only implementing fragment linking and fragment merging strategies, but also exhibiting fragment growing-based progressive exploration of uncharted chemical space through evolutionary trajectory analysis of fragment library development.

The evolutionary trajectory of the fragment library targeting PLpro (Fig. 8a) demonstrates successful incorporation of a critical pyridine ring structure through multiple rounds of fragment recombination and decomposition (with Fragment 1 and Fragment 2 first integrated into the library at the 5th and 10th iterations, respectively). Detailed binding site analysis confirms this structural evolution: 1) Fragment 2 completely maintains the $\pi-\pi$ stacking interaction pattern with TYR-268 (Fig. 8e), showing high conformational similarity to the original fragment (RMSD = \qty{1.1}{\angstrom}); 2) the newly introduced pyridine ring forms strong $\pi-\pi$ stacking interactions with TYR-268 (centroid distance = \qty{5.1}{\angstrom}, dihedral angle = \qty{86.1}{\degree}), resulting in a significant reduction of binding free energy (${\Delta\mathit{G}}=$ $-1.1$ $kcal/mol$). These findings conclusively validate the affinity enhancement achieved by AutoLeadDesign.
Study targeting the PRMT5 similarly reveals characteristic fragment evolutionary patterns (Figs. 8b, f-h). This fragment evolution process strictly adheres to the theoretical framework of fragment growing—a localized exploration paradigm that optimizes known fragments to achieve high-affinity derivatives. Furthermore, the affinity distributions of fragment libraries targeting PLpro (Fig. 8i) and PRMT5 (Fig. 8j) provide further evidence that the dual exploration of both uncharted and local fragment spaces systematically drives the fragment library updating toward desired pharmacological profiles.

\section{Methods}
\subsection{Overview of AutoLeadDesign}
Fig. 1 illustrates the workflow of AutoLeadDesign. Initially, molecules from the compound library are decomposed into structural fragments. These fragments are then filtered based on their docking scores to update the fragment library. Subsequently, key fragments are sampled according to their binding contributions to the target protein. These selected fragments are incorporated into a predefined prompt to guide a LLM in generating novel compounds. The newly designed molecules are added to the compound library for further iterative execution.

\subsection{Molecules decomposition}
The strategy employed in this study is based on BRICS (Breaking of Retrosynthetically Interesting Chemical Substructures), an advanced fragmentation approach derived from RECAP (Retrosynthetic Combinatorial Analysis Procedure), which introduced the concept of fragment space. This methodology typically utilizes 16 predefined rules to identify cleavable bonds within molecular structures. To prevent the generation of excessively small fragments with insufficient structural significance, we implement a minimum size requirement, mandating that each fragment contains at least two heavy atoms to ensure the extraction of meaningful chemical information.
\subsection{Fragment library update}
The fragment library at the t-th iteration comprises fragments decomposed from all previously designed compounds throughout the preceding k-1 iterations, resulting in a comprehensive yet redundant collection. Following the principle that ``fragments with lower affinity generally have reduced probability of recombining into high-affinity compounds``, we established a filtered library containing the top-k highest-affinity fragments selected from the global fragment pool. Furthermore, to quantitatively evaluate fragment significance, we assessed their binding contributions to target protein by calculating the average docking scores of all compounds containing each specific fragment. The score is calculated as:
\begin{equation}
	Score(f)=\frac{\sum_{s\in C,f\in Decompose(s)}{Docking(s,p)}}{|C|},
\end{equation}
where, $C$ stands for the compounds library containing all previously designed compounds until this iteration, and $Decompose(s)$ stands for the fragment set decomposed from compound $s$, and $Docking(s,p)$ stands for the binding energy of the compound $s$ targeting the protein $p$.
The sampling weight is caculated based on the scores:
\begin{equation}
	Weight(f)=\frac{Score(f)}{\sum_{f_{i}\in L}{Score(f_{i})}},
\end{equation}
where,  $L$ stands for the filtered fragment library. Through systematical sampling, AutoLeadDesign effectively extracts heuristic knowledge of known chemical space to guide LLM in further exploration of the uncharted chemical space.
\subsection{Molecules recombination}
The large language model (LLM) employed in our study is DeepSeek-v3, utilizing the ``deepseek-chat`` conversational model with a temperature parameter set to 1.5.
We have developed a standardized prompt template that, when integrated with sampled molecular fragments, guides the LLM to effectively explore the chemical space. Typically, the template is presented as follows:
\begin{quote}
	Generate a novel valid molecule SMILES which contains one fragment of [{$sample[0]$}, {$sample[1]$}, {$sample[2]$}] at least and do not generate any English text.
\end{quote}
 where $sample[0]$, $sample[1]$ and $sample[2]$ will be filled with the SMILES of sampled fragments.
\subsection{Dataset setup}
\noindent\textbf{Random initial compounds library.} To initialize the AutoLeadDesign pipeline, we randomly sampled 100 compounds from the ZINC100K database, and determined their protein binding affinities through docking simulations, providing the foundational start points for subsequent design cycles.

\noindent\textbf{Target protein.} We randomly select 10 target proteins in CrossDocked2022 database\cite{crossdocked}. These proteins belong to distinct classes: hydrolase, transferase, oxidoreductase, member protein and immune system protein, the PDB IDs are 4AAW, 14GS, 1FMC, 2AZY, 1PHK, 3U5Y, 5NGZ, 1E8H, 5TGN and 3HY9. In the case study, we chose two therapeutically significant targets: PRMT5 (PDB ID: 8UOB) and PLpro(PDB ID: 7L1G), both representing important clinical targets in their respective disease areas.

\noindent\textbf{Docking simulation.} 
We employed smina (a fork of Autodock Vina focused on improving scoring functions and energy minimisation) to approximate the Gibbs free energy changes during ligand-receptor binding\cite{smina}. Prior to docking, all ligands were converted from SMILES to 3D conformations using RDKit (v2023.9.6) with the ETKDG algorithm to ensure proper stereochemistry. These prepared conformations were then subjected to docking simulations using smina, with the binding site defined by the co-crystallized ligand coordinates with a 1 Å grid box expansion to ensure complete binding pocket coverage. Finally, the highest-affinity protein-ligand complexes were subjected to detailed binding interaction analysis in Maestro v14.2.

\subsection{The settings of compared methods}
\noindent\textbf{REINVENT\cite{reinvent}.} 
Building on policy-based reinforcement learning, REINVENT adapts recurrent neural networks for de novo molecular design task. The framework's innovative use of augmented episodic likelihood enables the directed exploration of chemical space, deigning compounds that satisfy predefined property requirements.

\noindent\textbf{ChemGE\cite{smilesga}.} ChemGE implements a population-based grammatical evolution strategy, where molecular structures undergo iterative optimization through evolutionary operations on their grammatical representations, facilitating directed exploration of chemical space..

Above baselines can be run using the software ($https://github.com/wenhao-gao/mol\_opt$) in practical molecular optimization benchmark\cite{molopt}.

\noindent\textbf{RGA\cite{reinforega}.}
RGA attempts to reformulate an evolutionary process as a Markov decision process and uses  an E(3)-equivariant neural network to choose parents and mutation types based on the 3D structure of the ligands and proteins. The model weight was obtained from their public repository ($https://github.com/futianfan/reinforced-genetic-algorithm$).

\noindent\textbf{LMLF\cite{46}.}
LMLF  separate the prompt into domain-constraints that can be written in a standard logical form, and a simple text-based query. The the constraints are
progressively refined using a logical notion of generalization to progressively design compounds. The settings of LMLF can be loaded from their repository ($https://github.com/Shreyas-Bhat/LMLF$).

Notably, to standardize the optimization objectives across all the frameworks, we unified the molecular property calculation module to focus exclusively on binding affinity with the target protein which is calculated by smina.

\subsection{Visualization}
\noindent\textbf{Clustering analysis.}To visualize and compare the chemical space of the initial and designed compounds, we constructed t-SNE plots. The 2048-bit Morgan fingerprints were used as molecular descriptors. Employing the Barnes-Hut implementation of the t-SNE algorithm, we obtained two-dimensional representations for clustering analysis.

\noindent\textbf{2D ligand-residue interaction.} For detailed analysis of compound-target protein interactions, the highest-affinity molecular conformation selected by SMINA docking was further analyzed using Maestro v14.2. Typically, hydrogen bonds were identified with a maximum bond length of \qty{2.8}{\angstrom}, requiring minimum angles of \qty{120}{\degree} for donors (D-H$\rightarrow$A) and \qty{90}{\degree} for acceptors (H$\rightarrow$A-X); halogen bonds were detected with a \qty{3.5}{\angstrom} distance cutoff and similar angular constraints of $\geq$\qty{120}{\degree} for donors (R-X$\rightarrow$A) and $\geq$ \qty{90}{\degree} for acceptors (X$\rightarrow$A-Y); $\pi-\pi$ interactions were classified as parallel stacking when within \qty{4.4}{\angstrom} with $\leq$ \qty{30}{\degree} angular deviation or as T-shaped stacking when within \qty{5.5}{\angstrom} with $\geq$\qty{60}{\degree} angles; while $\pi-cation$ interactions were considered significant within \qty{6.6}{\angstrom} and maximum \qty{30}{\degree} angular deviation. These carefully optimized parameters ensured comprehensive and accurate characterization of all potential molecular interactions at the binding interface.

\noindent\textbf{3D ligand-pocket binding pose.} Structural visualization of the ligand-protein complex was performed using PyMOL v3.1 (Schrödinger, LLC), while all intermolecular interactions were computationally characterized using Maestro.

\section{Conclusion}
In this work, we propose AutoLeadDesign, a novel property-conditioned molecular design algorithm that inspires large language models (LLMs) through molecular fragments while leveraging extensive domain knowledge embedded in LLMs to systematically explore fragment-based chemical space. Our approach can automatically design compounds demonstrating high affinity and novel interactions to target proteins without reliance on the difficult-to-acquire bioactive ligands.

With much higher docking score, AutoLeadDesign achieves better performance against the SOTA methods. We further valid AutoLeadDesign in lead compounds design task against PRMT5 and PLpro respectively. The designed compounds demonstrate lower binding energy than known inhibitors designed by experts and the binding site have been verified to be effective. Notably, we find AutoLeadDesign can design compounds demonstrating novel and valid binding mode, illustrating the ability of exploration rather than imitate the reference compounds, which is crucial in designing lead compounds to new targets that leak known inhibitors. We further trace the process of design and find that the workflow shares analogous mechanisms with FBDD, demonstrating clearly the process of fragment merging, fragment linking and fragment growing, which further explain why it works.

AutoLeadDesign demonstrates robust capability in generating bioactive compounds against target proteins, highlighting its significant potential for de novo drug design. Especially in the scenario lacking known bioactive ligands, AutoLeadDesign can still design compounds with valid binding interactions to target protein. Such unique capability positions AutoLeadDesign as an indispensable solution for high-stakes drug design endeavors, notably advancing Orphan Drug development against rare diseases.

\section{Data availability}
The code and all the data including initial compounds, target proteins and results are available at https://github.com/TH1052638958/AutoLeadDesign.

\bibliography{sn-bibliography}

\includepdf[pages=1]{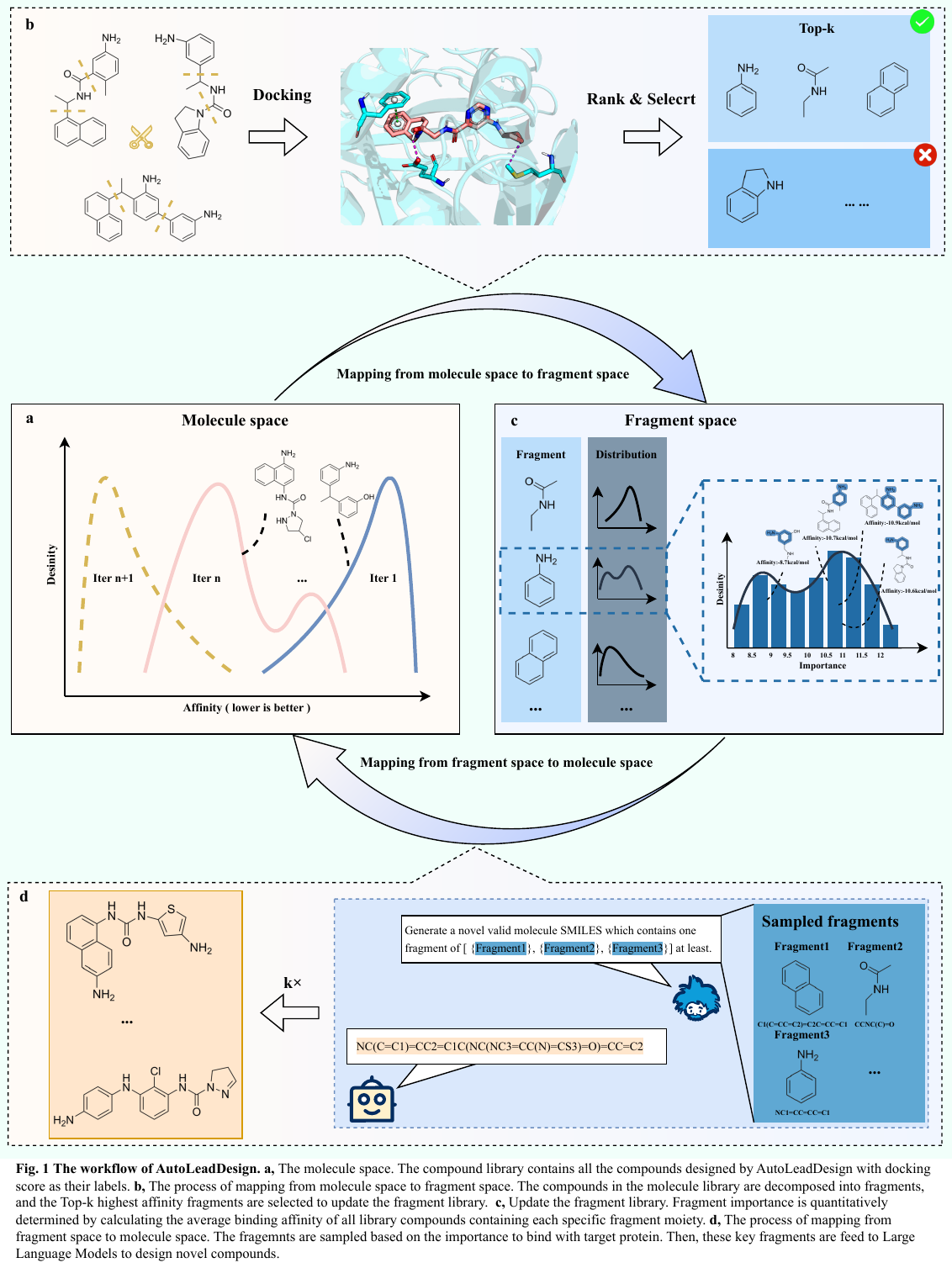}
\includepdf[pages=1]{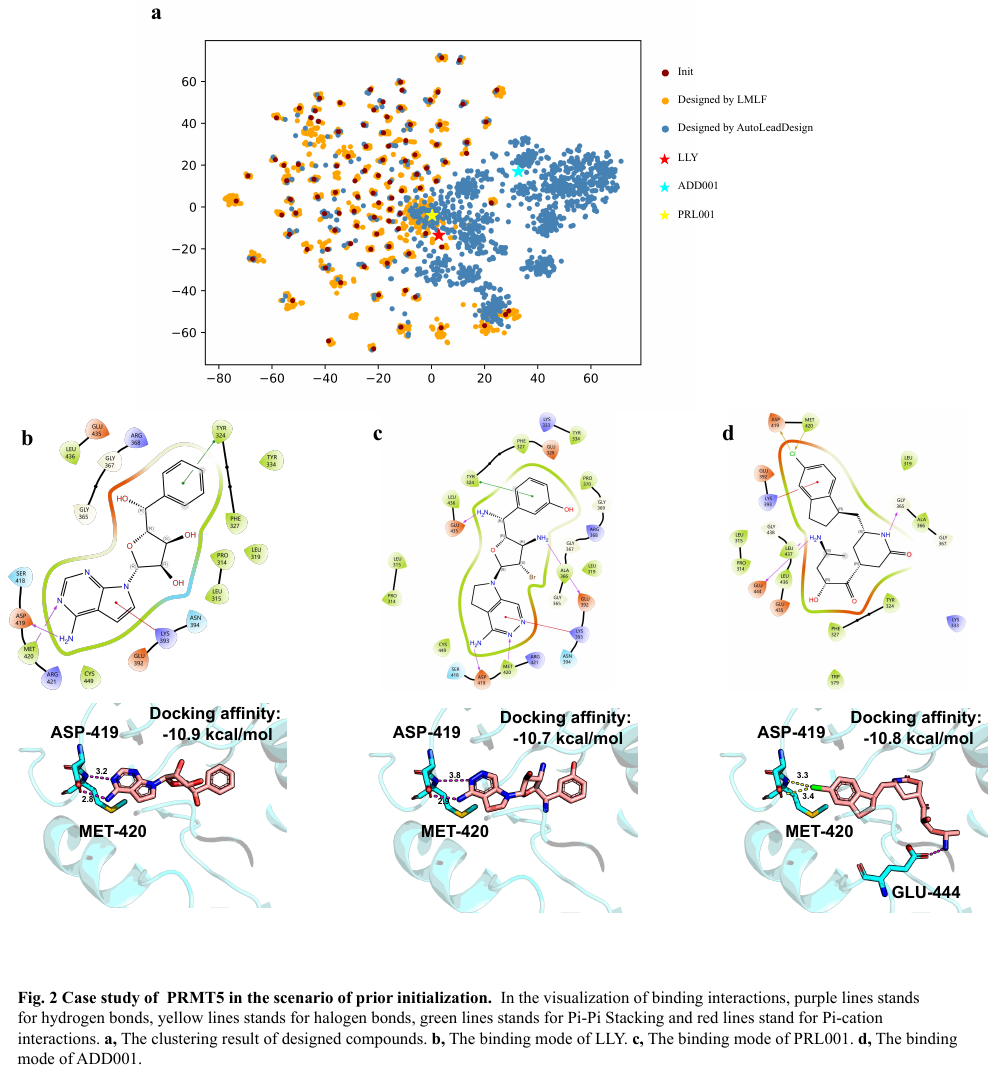}
\includepdf[pages=1]{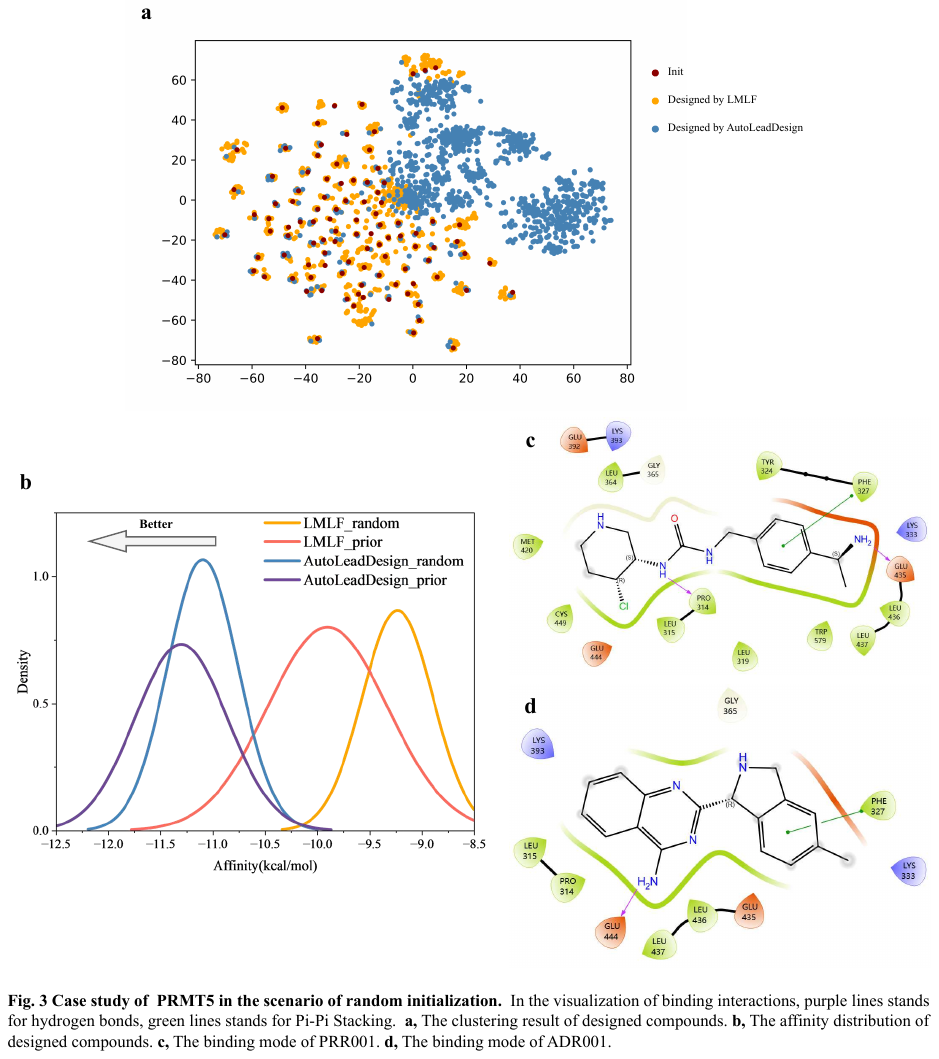}
\includepdf[pages=1]{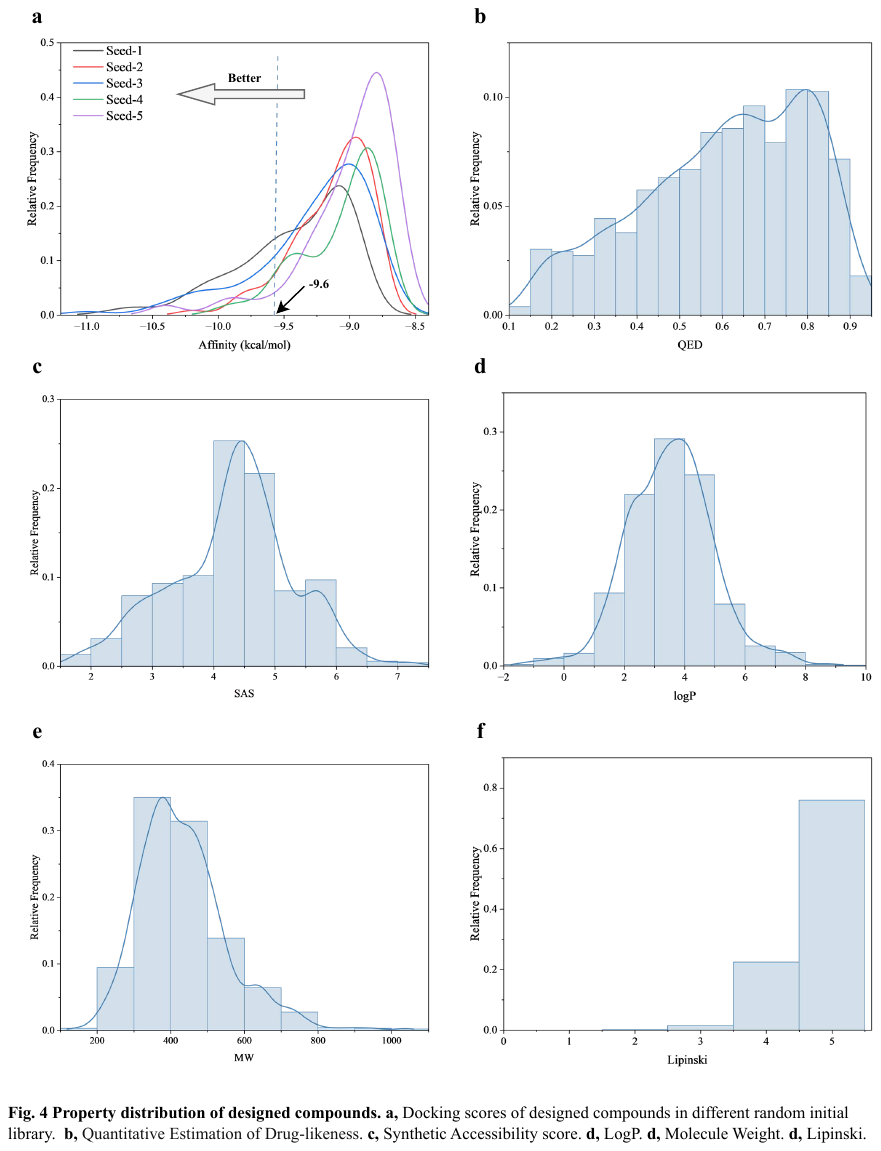}
\includepdf[pages=1]{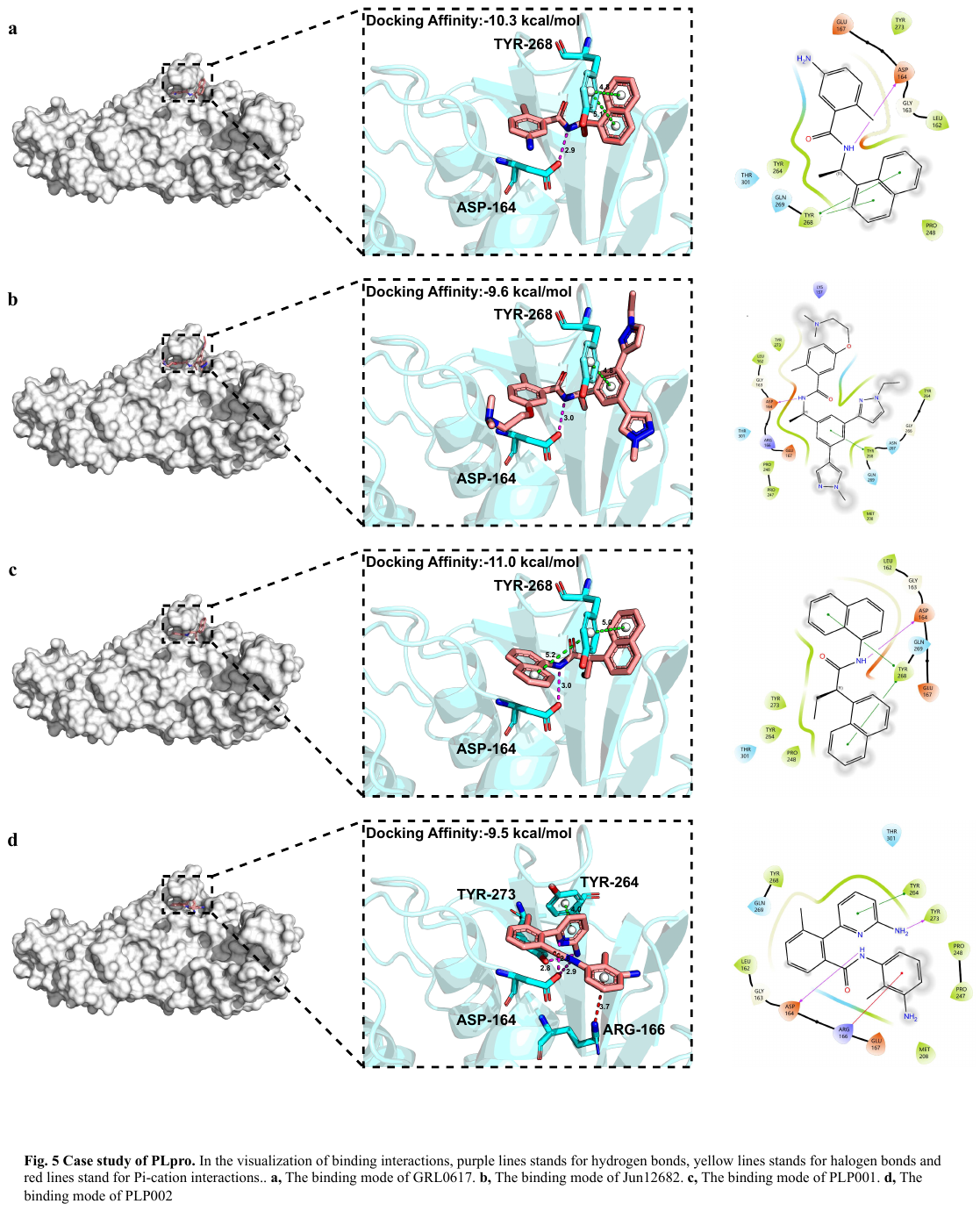}
\includepdf[pages=1]{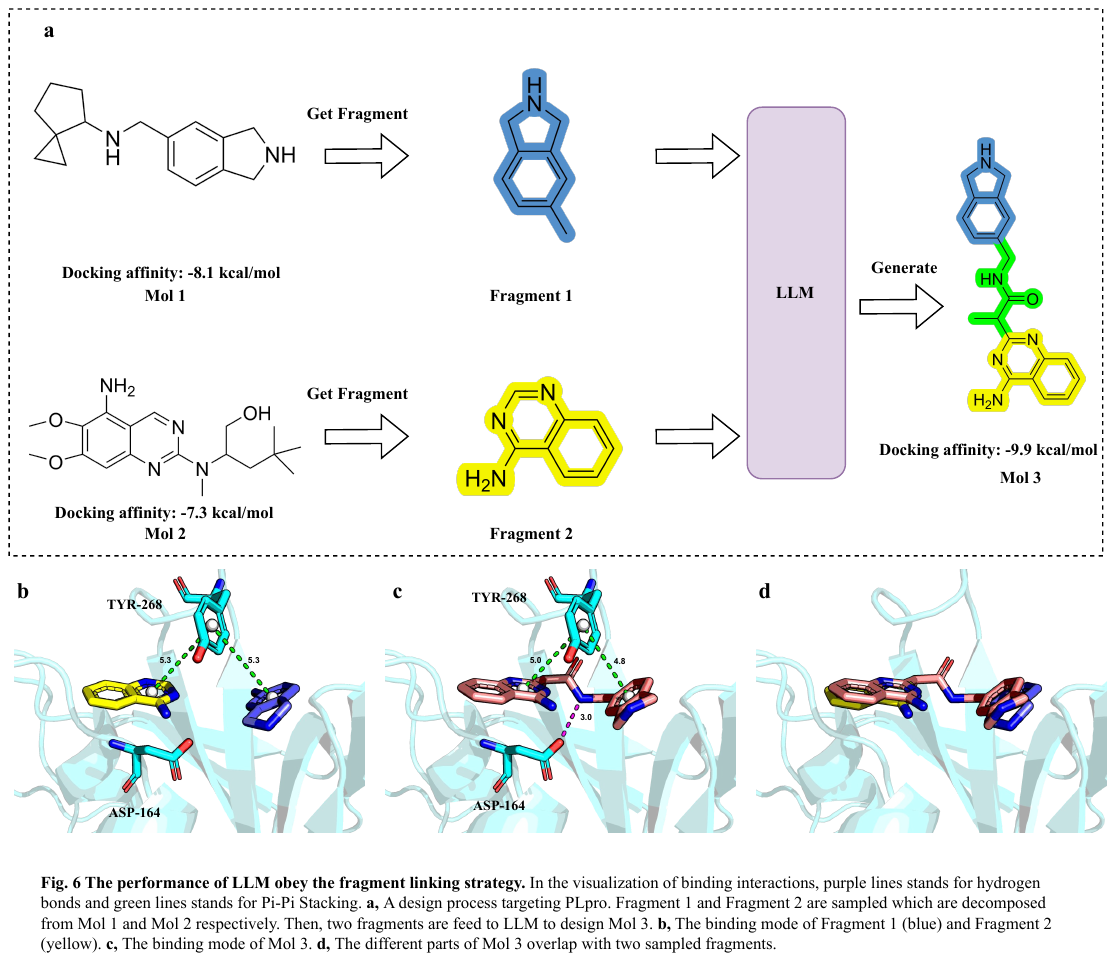}
\includepdf[pages=1]{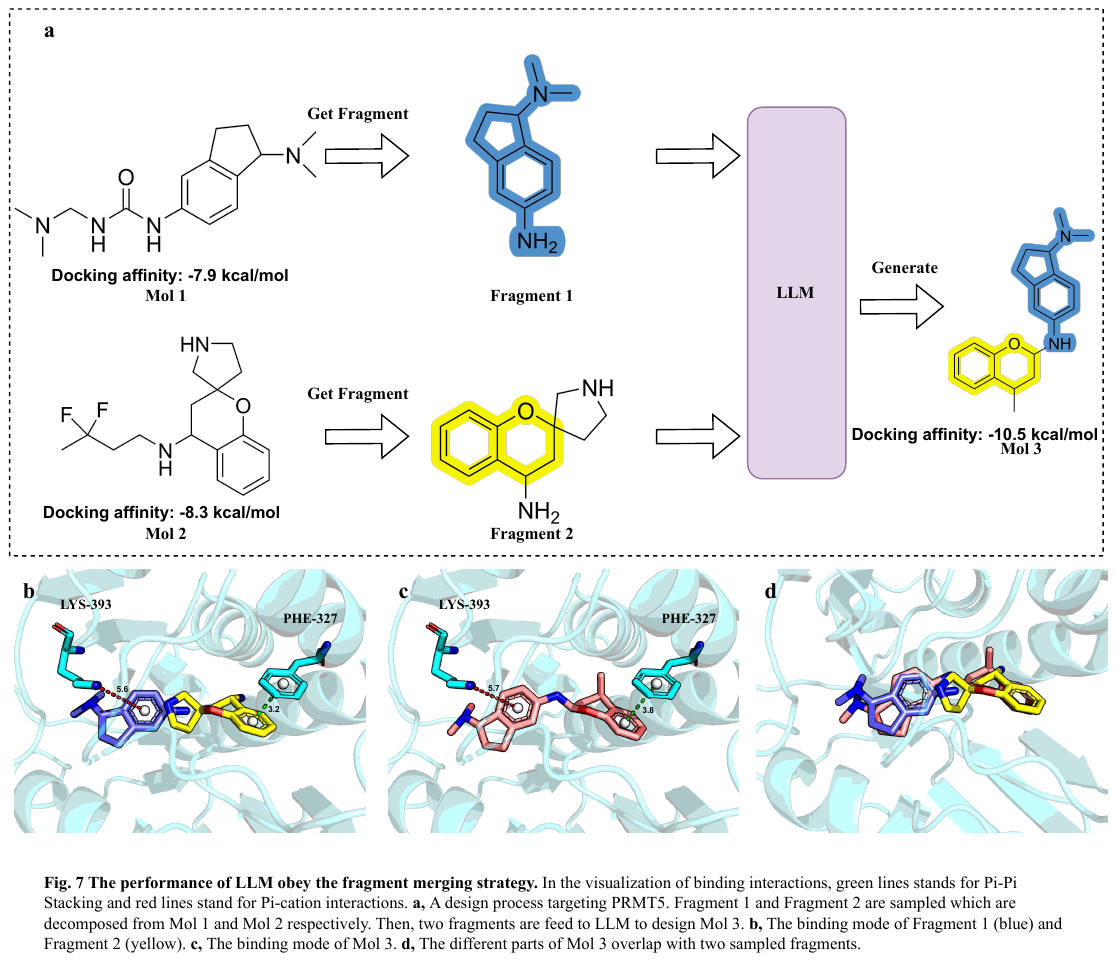}
\includepdf[pages=1]{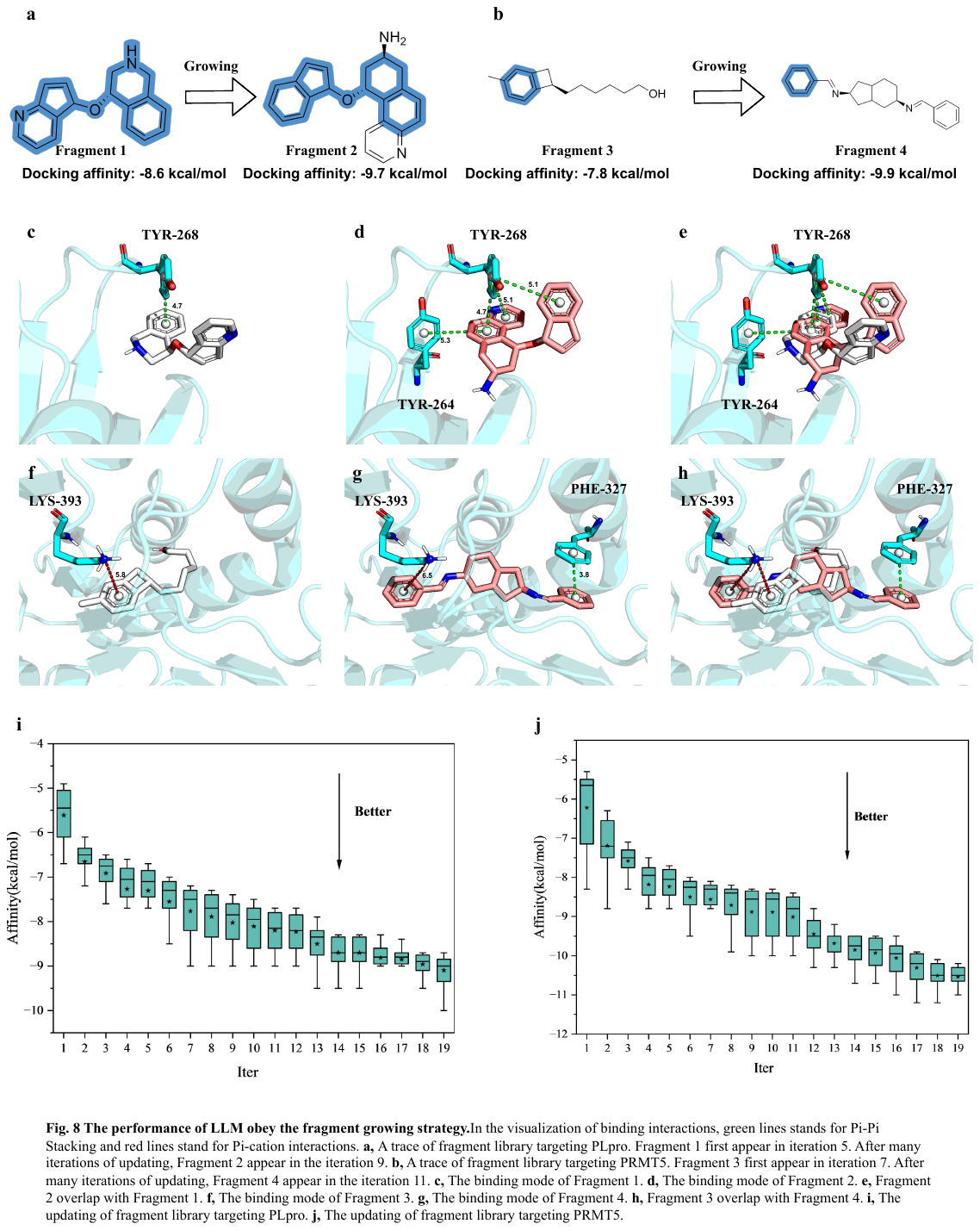}

\end{document}